# AI and Generative AI Transforming Disaster Management: A Survey of Damage Assessment and Response Techniques

Aman Raj
amanraj@google.com

Lakshit Arora
lakshit@google.com

Sanjay Surendranath Girija
sanjaysg@google.com

Ankit Shetgaonkar
ankiit@google.com

Dipen Pradhan
dipenp@google.com

Shashank Kapoor
shashankkapoor@google.com

Google

*Abstract*—Natural disasters, including earthquakes, wildfires and cyclones, bear a huge risk on human lives as well as infrastructure assets. An effective response to disaster depends on the ability to rapidly and efficiently assess the intensity of damage. Artificial Intelligence (AI) and Generative Artificial Intelligence (GenAI) presents a breakthrough solution, capable of combining knowledge from multiple types and sources of data, simulating realistic scenarios of disaster, and identifying emerging trends at a speed previously unimaginable. In this paper, we present a comprehensive review on the prospects of AI and GenAI in damage assessment for various natural disasters, highlighting both its strengths and limitations. We talk about its application to multimodal data such as text, image, video, and audio, and also cover major issues of data privacy, security, and ethical use of the technology during crises. The paper also recognizes the threat of Generative AI misuse, in the form of dissemination of misinformation and for adversarial attacks. Finally, we outline avenues of future research, emphasizing the need for secure, reliable, and ethical Generative AI systems for disaster management in general. We believe that this work represents the first comprehensive survey of Gen-AI techniques being used in the field of Disaster Assessment and Response.

## I. INTRODUCTION

The combination of growing urban areas and climate change is leading to more natural disasters and resulting in devastating human and economic losses. Precise and timely damage assessment is one of the most significant factors in an effective disaster response. This calls for innovation in disaster management and assessment techniques. It is required for efficient resource allocation, rescue planning, and reconstruction planning. Without clear and comprehensive damage assessment, the response is delayed and ineffective.

Traditionally, damage assessment has relied on manual methods like ground survey, satellite data interpretation and physical inspection which are valuable, but nevertheless slow and labor-intensive [1]. They are also hindered by communication failures and other logistic issues. AI/GenAI methods like deep learning models offer a revolutionary solution for disaster damage assessment. Models like Variational Autoencoders (VAEs) [2] and Generative Adversarial Networks (GANs) [3], can synthesize representations of disaster-affected areas, enabling rapid analysis and simulation. Generative AI models, particularly task-finetuned ones, also are very well-suited to perform image recognition and classification on multi-modal data from affected regions, greatly improving the speed of assessment [4].

In this paper, we conduct a thorough evaluation of AI and GenAI's capabilities in assessing damage from disasters like earthquakes, wildfires and cyclones, with a focus on their strengths and limitations. We further discuss its capacity to integrate disparate data across various modalities, including satellite imagery, social media posts, sensor networks, and emergency reports, to generate real-time situational awareness. Towards the end we provide an overview, highlighting the most important challenges and propose potential directions for future research and development in this critical area.

The organization of this paper is as follows. In section II, we discuss the various domain-specific literatures on the use of GenAI in disaster assessment. In section III, we discuss the data modalities across which GenAI techniques can be utilized. In section IV, we present how Gen-AI techniques can also be used for combating misinformation and in an adversarial setting. Section V brings the discussion on data privacy, security considerations and deployment challenges and section VI is about our opinion on the future for Gen-AI techniques and conclusions.

## II. DOMAIN-SPECIFIC LITERATURE REVIEW

This section provides a review of recent publications on damage assessment techniques for earthquakes, wildfires, and cyclones with a focus on the use of AI/GenAI.

### A. Earthquakes

Earthquakes present critical challenges to damage assessment due to the widespread and often unpredictable extent of damage they inflict. Traditional earthquake damage assessment has relied heavily on visual inspection of structures and infrastructure by professional engineers and emergency teams [1]. These procedures, while vital in determining structural soundness and identifying imminent safety hazards, are time-consuming and often risky. Remote sensing technologies, satellite high-resolution images and information from unmanned aerial vehicles (UAV, or drones) are increasingly being used along with machine learning to automatically identify and distinguish damaged structures and infrastructure parts [5, 6].

Lately there has been significant research on using GenAI techniques for the damage assessment. Mousavi, S. Mostafa, et al. (2024) explored the use of multimodal social media data to estimate earthquake shaking intensity [7]. Their work leveraged Natural Language Processing (NLP) techniques, such as sentiment analysis using transformer-based models, on text data combined with image classification of user-submitted photos to assess damage levels in real-time. They addressed the challenge of noisy social media data by employing filtering techniques based on source credibility and content verification. Estêvão (2024) evaluated multimodal Generative AI models (like OpenAI's GPT-4o and GPT-4o mini, and Google's Gemini 1.5 Pro and Gemini 1.5 Flash) for classifying building damage post-earthquake from images using the EMS-98 scale [59]. Their study found varying accuracy (28.6%-75.0%) across masonry and reinforced concrete structures, best accuracy shown by GPT-4o. The work showed GenAI has the potential for automated assessment but indicated that current performance isn't optimal, suggesting significant improvements could be achieved via fine-tuning or Retrieval-Augmented Generation (RAG) [8].

Bhadauria (2024) conducted a comprehensive review exploring the blend of AI and Machine Learning (ML) tools within earthquake engineering, focusing on damage assessment and retrofitting strategies. Their work surveyed recent applications of ML, Pattern Recognition (PR), and Deep Learning (DL), emphasizing how these techniques surpass traditional model limitations in areas like risk prediction, retrofitting design, and structural optimization. The review identified challenges, limitations, and future research avenues for advancing seismic resilience engineering using AI/ML [9].

*B. Wildfires*

Wildfire damage assessment is determining burn severity and comprehending the total impact on infrastructure, ecosystems, and communities. Classic approaches in the past employed remote sensing imagery from optical, thermal, and radar sensors [10, 11, 12]. Generative AI is introducing new paradigms for analyzing and interpreting such information.

Asanjan et al. (IGARSS 2024) proposed a technique for wildfire segmentation from remotely sensed data using a quantum-compatible, modified Vector-Quantized Variational-Autoencoder (VQ-VAE) [13]. Their methodology enhanced probabilistic U-Net [57] models by incorporating a Restricted Boltzmann Machine (RBM) [58], which is optimized with Parallel Tempering in PySA, into the prior encoder step for improving latent space description and segmentation representation [14]. This approach helps to leverage its unique probabilistic nature for a more expressive latent representation, aiming to better model the complex patterns in wildfire data compared to purely classical approaches. Using the pre-fire image as condition, the VQ-VAE decoder generates a reconstruction of the post fire image. The difference between this generated image and the actual post-fire image is processed to produce a segmentation map of the burned area and is then evaluated using metrics like Intersection over Union (IoU) and F1-Score.

Park et al. (2020) introduced a deep learning wildfire detection system using remote camera images with the objective of solving data imbalance issues typically encountered in disaster datasets [15]. They used CycleGAN [16], a particular GAN for unpaired image-to-image translation, mostly to generate synthetic wildfire images from accessible non-wildfire forest photographs. Such data augmentation based on CycleGAN increased training data numbers and variety widely without employing paired examples to overcome the constraint of limited wildfire imagery. Then, a DenseNet-driven classifier was trained on the data-augmented set and performed better (98.27% accuracy and 98.16 F1-score) than those with no GAN augmentation or by utilizing traditional pre-trained models like VGG-16 [60] and ResNet-50 [61]. The research confirmed that GAN-generated synthetic data were able to efficiently enhance wildfire detection accuracy.

GenAI's capacity for synthetic training data generation addresses one of the most important issues of wildfire estimation: the unavailability of labeled data. Techniques like CycleGANs can be used to transform pre-fire to post-fire scenes, synthesizing realistic training examples for damage segmentation models.

*C. Cyclones*

Cyclone damage estimation demands prompt and correct quantification of storm impacts along coastlines, structures, and facilities to guide effective disaster management [17]. Traditional methods for this were based on the combination of satellite imagery, aerial surveys, and ground observation data [18]. Modern AI offers advanced alternatives for automating and streamlining such tasks.

Arachchige & Pradhan (2025) developed a machine learning-based risk estimation of hurricane-induced flood damage in Florida from granular insurance data enriched with remote sensing variables [19]. They compared several algorithms (Random Forest [62], XGBoost [63], GBM, Neural Networks) and selected a stacked ensemble model, which combines predictions from these base learners by a meta-learner, because of its enhanced performance. This ensemble model predicted flood damage at the Zip Code Tabulation Area (ZCTA) level with a mean absolute error of 11.3%. Explainability tools (Partial Dependence Plot [64]) were used for model interpretation, identifying key predictors like property value and showing geographic hotspots based on factors like Gulf warmth and urbanization.

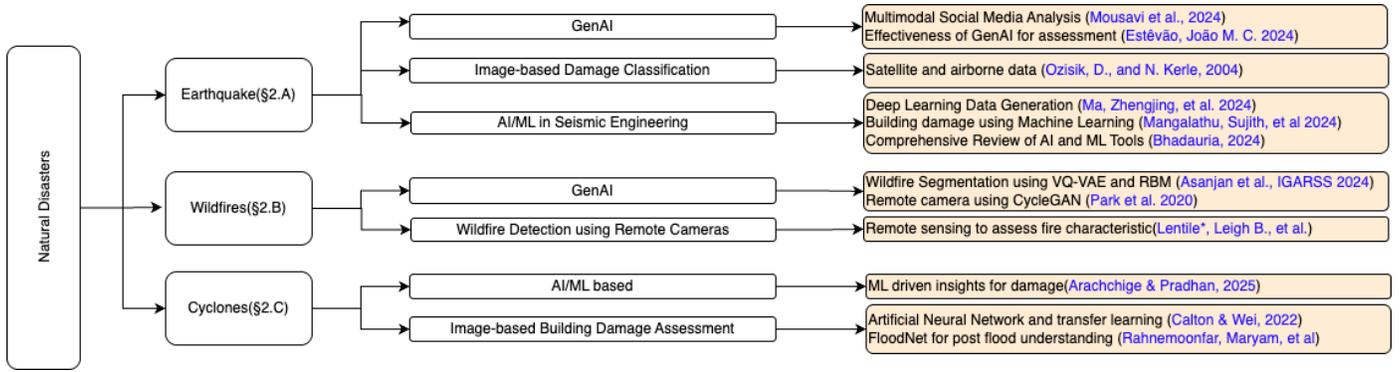

Figure 1: Taxonomy of Natural Disaster surveys and methods in this survey

Calton & Wei (2022) utilized transfer learning on advanced Artificial Neural Networks (ResNet, MobileNet [65], EfficientNet [66]) to assess hurricane/cyclone building damage from images [20]. Their work involved binary classification (flood/non-flood) and multi-class object detection (damaged roof, wall, flood, structural damage). MobileNet demonstrated superior performance, achieving 87% accuracy (0.88 F1-score) in classification and high confidence (up to 97.58%) in detecting specific damage types, showing the efficiency of transfer learning for rapid post-cyclone assessment.

In addition, GenAI can assess infrastructure resilience by generating hypothetical "damaged" copies of critical structures [21]. For instance, it is possible to train a VAE to encode the structural response of bridges subjected to various simulated storm loads. The encoder maps the pre-cyclone bridge state into a latent space, and the decoder generates potential post-cyclone states based on estimated damage parameters. This enables predictive flagging of vulnerable infrastructure and informing targeted inspections following the storm [21].

### III. DATA MODALITIES AND GENERATIVE AI APPLICATIONS

This section describes GenAI's applications for damage assessment and disaster response across various data modalities.

#### A. Text

Text data offers dense, critical information sourced from various sources like social media posts, news services, emergency notification services, call center reports, and government communication released following a disaster. Text data analysis is of paramount importance due to its natural language characteristics, thus enabling it to convey information, express emotions and reflect human communication patterns. Generative AI techniques are revolutionizing the use of this text data in disaster relief.

Instead of relying on conventional sentiment analysis, there are now newer approaches that employ transformer-based models such as BERT, RoBERTa, and XLM-RoBERTa [22]. Training or finetuning the models against multilingual, disaster-oriented corpora allows public opinion related to developing events, resource constraints, and emotional reactions in affected populations to be better identified. Furthermore, aspect-based sentiment analysis is used to classify fine-grained sentiment on specific entities, like infrastructure or government response, further improving resource targeting accuracy. Tarasconi, Francesco, et al(2017) published research suggesting that GenAI also supports relationship extraction and event correlation resolution beyond mere Named Entity Recognition [23]. Models can identify complex relationships among damaged infrastructure, affected populations, and relief responses, providing richer situational context than traditional rule-based systems.

Models are able to leverage unstructured information and contextual knowledge in order to deduce top damage indicators, estimate affected populations, and indicate emergent needs, making reports tailored to the specific needs of the individual stakeholder, either policymakers or first responders. Smart Conversation Agents (Chatbots) no longer stop at responding to standard questions. GenAI makes adaptive dialog management and contextual responses possible. Through a mix of reinforcement learning and human feedback (RLHF), chatbots can be optimized for more empathetic, contextually knowledgeable interactions, creating greater user trust and encouraging active engagement with usable resources [24].

Managing the inherent subtleties of text data requires sophisticated methods. Text data from social media can be noisy and unreliable, and requires filtering based on the source for reliability [28]. Another possible issue is disinformation and combating it requires GenAI to be able to recognize and mark-off manipulated stories, integrate fact-checking modules, and actively recommend reliable sources of information [29]. When dealing with multilingual sources of data, accurate machine translation and cross-lingual information searching are required.

#### B. Image Data

Image data from satellites, aerial drones, roadside cameras, and social media sites is very important for visual and quantitative disaster damage assessments [30].

Automated damage detection and segmentation now employ advanced Convolutional Neural Networks (CNNs) like Mask R-CNN and U-Net, implementing semantic segmentation for pixel-level classification of specific damage types [31]. To improve robustness against variations in illumination, weather, and image quality, transfer learning is used by pretraining over large datasets like ImageNet and then fine-tuning with disaster-related datasets like xBD. Furthermore, techniques like attention mechanisms and spatial pyramid pooling [50] extract multi-scale contextual information [32], facilitating precise damage extraction even in partially obscured scenes.

When image data is incomplete, image inpainting techniques are invaluable. Generative models, particularly those based on contextual autoencoders and Generative Adversarial Networks (GANs), enable realistic restoration of corrupted or obscured regions [33]. The generator network learns feature representations from undamaged areas to synthesize realistic in-paintings, while the discriminator enforces realism, guided by adversarial loss functions. Image super-resolution uses architectures like Enhanced Deep Super-Resolution Network (EDSR) [67] and Residual Channel Attention Networks (RCAN) [68] to enhance low-resolution images. This uses residual blocks and channel attention to reconstruct high-frequency details and sharpen damage maps [34].

GANs and VAEs can produce new images simulating various levels of structural damages and weather conditions addressing the scarcity of labeled training data. They significantly improve the strength and generalization capabilities of AI algorithms to effectively recognize damage patterns in new scenarios. The key challenges in this area include accommodating variations in image resolution and quality, and attaining high levels of precision for GenAI-driven damage assessments.

*C. Video Data*

Video data creates a stream of information that is critical for facilitating rapid damage assessment. Video maintains event chronology, so it is possible to place incidents in a more nuanced context to estimate the intensity and progression of damage. Drone videos, security camera footage, news coverage, and live streaming offer various insights and GenAI techniques can help uncover actionable insights from this dynamic information.

AI significantly enhances real-time damage assessment capabilities through enabling classification and identification of various forms of damage as and when they occur in video streams [35]. Spatiotemporal feature extraction with 3D Convolutional Neural Networks (3D-CNNs) and temporal dependency with Recurrent Neural Networks (RNNs) like LSTMs can track the general progression of a disaster event in real time. For instance, object detection software based on standalone frames can identify collapses of buildings or flooding water to issue evacuation warnings and guide rescue forces [36]. The feedback loop in real-time increases situation awareness so that resources are moved faster.

GenAI also makes it possible to identify events like collapses of buildings or explosions in video and trigger alerts. Object detection software like YOLO [69], enriched with temporal reasoning, can categorize such incidents swiftly, enabling quick damage evaluation and response. Moreover, GenAI is capable of aiding 3D reconstruction of damaged areas from video. Techniques like Structure from Motion (SfM) and Multi-View Stereo (MVS), complemented with deep learning for enhanced feature matching and depth estimation, can create high-resolution 3D models [37]. These models can provide structural strength estimation, rubble volume estimation, and can be combined with virtual reality packages for remote inspection.

Efficient use of video data has some open challenges. The massive volumes of data produced by video streams demand powerful computing and efficient processing techniques. The removal of motion blurring effects, occlusion, illumination variation, and view variation to maintain analysis quality is crucial. Finally, the privacy of the people captured in the video data must be maintained by anonymization techniques and adherence to ethical standards [38]. Research has to overcome these challenges to utilize the full potential of video data in disaster response.

*D. Audio Data*

Audio data like emergency calls, social media audio and recordings, and drone audio recordings hold crucial information on disaster damage like location, affected population size and impact severity. GenAI opens new possibilities for tapping into this data to enhance disaster response efforts.

Distress call analysis employs GenAI coupled with signal processing to identify and prioritize the most critical emergency situations [39]. Techniques like Mel-Frequency Cepstral Coefficients (MFCCs) [70] extract vocal features, but GenAI's strength lies in using models like transformers to understand call context beyond keywords. GenAI can be used to identify panic or distress and generate a more nuanced "urgency score", factoring in emotional tone and acoustic environment.

Sound event detection also benefits from GenAI. While Hidden Markov Models (HMMs) were traditional, deep learning models now learn complex acoustic patterns to detect events [40]. Another key advance through GenAI is the generation of synthetic audio data of disaster sounds (e.g., building collapse). This augments training and boosts detection accuracy, especially for rare events. Correlating these events with other modalities can also increase efficacy [41].

The key challenges with audio data are removal of background noise, accents, and maintaining privacy. While conventional signal processing tools help, GenAI can potentially generate sanitized audio samples or modify voices to restrict accent bias and ensure data anonymization. Future solutions could rely on GenAI to not just analyze audio data but generate insights and improve data quality.

IV. MISINFORMATION AND ADVERSARIAL ATTACKS

The ability to generate very realistic but entirely fake images and videos is a genuine threat to trustworthy GenAI-based assessment. Manipulated media ingested as authentic data can lead to flawed analysis and can disrupt the pipeline, leading to resource misallocation and even lethal consequences.

One of the biggest issues is employing images produced by GANs and diffusion models. GANs can produce manipulated images by modifying textures, erasing or adding objects, or changing environmental factors. Diffusion models that learn to reverse a gradual process of noise, introduce even greater realism and control over image synthesis, thus producing bogus evidence of destruction.

CNNs can help with this by analyzing minute inconsistencies that are oblivious to the naked eye [42]. They look at lighting, shading, texture, and even frequency-domain artifacts that are introduced in GAN or diffusion model processing. More recent architectures such as Vision Transformers (ViTs) [71] and Swin-Transformers [72] also offer improved capabilities at capturing global context in images that assist in detecting subtle inconsistencies. Techniques like inspection of noise patterns can detect the characteristic "fingerprint" produced by different generative models. In order to specifically counter deep fakes, techniques evaluating facial micro-expressions and lip-syncing inconsistency are essential [43, 44]. Development of "universal detectors" which can identify manipulations in any generative source is a key research goal.

Effective mitigation should also involve prevention along with detection. Digital watermarking, for example, can embed invisible yet verifiable signatures into the original image for authentication [45, 55]. Cryptography hash functions [56] along with perceptual image hashing provide an additional layer of content verification. Designing GenAI assessment systems with internal attention mechanisms to evaluate input reliability is also essential [46]. Furthermore, strong data governance imparts intrinsic trust upon GenAI models by providing tangible assurances about the underlying training data. This involves technical procedures for data lineage tracking, automated quality validation, privacy enforcement minimizing re-identification, secure data lifecycle management, and bias auditing before training.

The assurance of integrity, security, and ethical handling of core data provides the necessary verifiable foundation for trust in the model's predictions during high-stakes events.

## V. Privacy, Security and Deployment Challenges

It is important to have strong data security and privacy safeguards when implementing GenAI for disaster response. Overlooking them can weaken public trust, risk sensitive information, and cause unintended harm.

Protecting people's privacy by having strong anonymization of data is critical. K-anonymity and l-diversity can be used to keep sensitive attributes from being easily traced back to individuals in the data set [47]. Differential Privacy (DP) also offers stronger, provable privacy guarantees [51]. The Laplace Mechanism, a core component of DP, adds calibrated noise to data in a bid to hide individual contributions [48]. DP allows for fine-grain privacy budget tuning for privacy versus model utility trade-offs in specific disaster assessment applications.

Secure data storage and transmission demand robust cryptographic controls. Penetration testing and regular security audits identify and correct GenAI system architecture weaknesses. Adopting zero trust architectures across all the data is a main idea to keep in mind for all such safety types.

Explainable AI (XAI) approaches can provide insight into model reasoning [49]. SHAP (SHapley Additive exPlanations) [73] importance quantifies the contribution of every input feature to a prediction, while LIME (Local Interpretable Model-agnostic Explanations) [74] approximates local model behavior with simpler, interpretable models that allow model output understanding via local explanations to help establish trust in stakeholders.

Possible biases of AI models should be addressed in order to prevent unfair outcomes. Right training data choice and pre-processing are necessary in order to present diverse populations and regulate possible biases [54].

Deployment concerns add complexity in terms of real-time computation, which is highly important within disaster zones, and calls for compression and optimization technology on models so they can be run effectively on all kinds of hardware.

## VI. Future Prospects and Conclusion

AI and GenAI ability to combine various data sources and types in new ways can help transform damage assessment. Unleashing this potential will require focused research work in priority areas.

**Advancing GenAI Model Capabilities:** This requires focused research on improving GenAI models for precision, generalizability, and resilience. It needs exploring the application of powerful deep learning architectures, like transformers with advanced attention and graph neural networks (GNNs) [75] for spatial relationships modeling. Research also needs to be performed to develop methods that can properly handle noisy and missing data, and to make the models more robust against adversarial attacks and false information using adversarial training and advanced data augmentation. Unified datasets and benchmarks are also needed to facilitate progress.

**Integration with existing technologies:** Creating integrated disaster response platforms also need integration of GenAI with other existing systems. This can be achieved by integration of GenAI with i) Geographic Information Systems (GIS) [76] for spatial analysis and visualization, ii) diverse sensor networks for real-time data ingestion, iii) with communication systems for disseminating actionable intelligence. Research should target efficient and scalable interfaces between these systems.

**Explainable AI (XAI):** Developing systems that explains the behavior and results produced by a model to improve transparency and trust in AI-driven insights [52]. This includes visualizing model activations, identifying influential features (e.g., via SHAP, LIME), and generating human-interpretable explanations, potentially using standardized formats like model cards.

**Addressing Ethical and Societal Issues:** Ethical considerations must guide GenAI design and deployment in disaster response [53]. This requires ethical frameworks for data handling, addressing privacy (e.g., via federated learning, DP) and security, mitigating biases through fairness-aware learning and representative data, and ensuring equitable resource distribution. Interdisciplinary collaboration among computer scientists, ethicists, and responders is essential to attain this goal.

**Developing Novel GenAI Applications:** GenAI enables advanced applications beyond basic assessment. Future work includes personalized disaster response plans using privacy-preserving methods [51], predictive resource allocation models forecasting needs, realistic evacuation simulations to optimize strategies, and automated damage report generation using Natural Language Generation (NLG) [77] to synthesize multimodal data into coherent summaries.

In conclusion, this survey shows that AI and GenAI provide a powerful instrument for transforming disaster damage assessment. Realizing its full potential depends on overcoming significant challenges in model robustness, data quality, ethics, and integration. There are concerns around ethical deployment, measures against disinformation, anonymization and adversarial attacks. The successful integration of strong ethical frameworks and reliable technical safety measures will ultimately determine the future effectiveness and societal acceptance of GenAI in disaster assessment and response.

We believe that this work represents the first comprehensive survey of GenAI techniques used for disaster assessment and response. In particular, we provide a deep overview of how this emerging field of technology can be used for damage assessment for earthquakes, wildfires and cyclones. The research conducted in this paper demonstrates the potential for the AI and GenAI in disaster response, and also the need for further research and investment in the domain to realize this potential.